\documentclass[useAMS,usenatbib,usegraphicx]{mn2e}

\title[Testing MOND with Local Group spiral galaxies]
{Testing MOND with Local Group spiral galaxies}
\author[E. Corbelli and P. Salucci]
{Edvige~Corbelli$^{1}$\thanks{E-mail: edvige@arcetri.astro.it} and
 Paolo~Salucci$^{2}$
\\
$^{1}$INAF-Osservatorio Astrofisico di Arcetri, Largo E. Fermi, 5
I--50125 Firenze, Italy
$^{2}$SISSA, Via Beirut 2-4, 34019 Trieste
I-Italy}

\begin{document}
\date{Accepted 2006 October 18. Received 2006 October 11; in original form 2006
June 14}
\pagerange{\pageref{firstpage}--\pageref{lastpage}} \pubyear{2006}
\maketitle
\label{firstpage}

\begin{abstract}

The rotation curves and the relative mass distributions of the two nearby 
Local Group spiral galaxies, M31 and M33, show discrepancies with  
Modified Newtonian dynamic (MOND) predictions. In M33 the discrepancy lies 
in the kinematics of the outermost regions. It can be alleviated 
by adopting tilted ring models compatible with the 21-cm datacube but 
different from the one that best fits the data. 
In M31 MOND fails to fit the falling part of the rotation curve at 
intermediate radii, before the curve flattens out in the outermost regions. 
Newtonian dynamics in a framework of a stellar disc embedded in a dark halo
can explain the complex rotation curve profiles of these two galaxies, while 
MOND has  
some difficulties. However, given the present uncertainties in the kinematics  
of these nearby galaxies, we cannot address the success or failure of MOND  
theory in a definite way. More sensitive and extended observations around 
the critical regions, suggested by MOND fits 
discussed in this paper, may lead to a definite conclusion.

\end{abstract}

\section{Introduction}

According to Newtonian dynamics the mass distribution of the luminous 
components of spiral galaxies cannot account for the observed profiles
of their rotation curves (hereafter RC) especially in the outer regions of 
galaxies (Rubin, Ford, $\&$ Thonnard 1980). To account for this discrepancy, 
which becomes more pronounced for late type galaxies, discs are thought to be
embedded in dark halos of non baryonic matter (Persic, Salucci  $\&$ Stel 1996). 
An alternative explanation for the mass discrepancy has been  proposed by Milgrom  
by means of the modified Newtonian dynamics or MOND (Milgrom 1983). 
According to this theory the dynamics becomes  
non-Newtonian below a limiting  acceleration value, ${a_0}\sim 10^{-8}$~cm~s$^{-2}$, 
where the effective gravitational acceleration takes the value 
$a_{eff} = \sqrt{a_0 g_n}$, with ${g_n}$ the acceleration in Newtonian dynamics.   
Outside the bulk of the mass distribution, MOND predicts a much slower decrease of  
the (effective) gravitational potential, with respect to the Newtonian case. This   
is often sufficient to explain the observed non-keplerian behavior of RC
(Sanders $\&$ McGaugh 2002, Lokas 2002). 

This success is remarkable in that MOND has only one free parameter, namely $a_0$. 
MOND theory of Gravity has been 
generalized in the framework of a relativistic, Lorentz invariant theory 
(Bekenstein 2004). It is worth noticing that MOND 
could account for the acoustic peak structure of Cosmic Microwave Background 
experiments (Slosar, Melchiorri, $\&$ Silk 2005, McGaugh 2004).
Gravitational lensing data together with X-ray data in galaxy cluster regions give 
perhaps the most convincing proof for the existence of  "dark matter"
(Clowe, Gonzalez $\&$ Markevitch 2004; 
Pointecouteau $\&$ Silk 2005) though on much larger scales than we consider here.  
On the other side, there are difficulties 
for the dark matter paradigm as well (e.g. Goerdt et al. 2006) and dark particles 
have not been detected yet.

It is however important to check MOND validity on galactic scales, that host    
the empirical phenomenon which stimulated its birth and its theoretical development.
MOND is unable to fit some RC of spiral galaxies (e.g. Gentile et al. 2004) 
but this failure is not decisive to totally rule out this  theory of Gravity. 
In fact  MOND mass modeling  depends on the exact value of the acceleration 
threshold  $a_0$, on  the galaxy distance  and  on the detailed distribution 
of stars and gas. Uncertainties in these quantities weaken the relevance of the
comparison between model predictions and data. All the above, in addition to the 
possible presence of bars, interactions and warps, makes the MOND-dark matter 
debate still open (Bottema et al. 2002).  In this perspective M31 and M33, well 
studied  spirals and primary distance indicators, are excellent benchmarks for 
MOND. For these galaxies, all the above uncertainties are fairly small: distances 
are well known (780 and 840~kpc within 10$\%$), and their exponential stellar discs 
and gas surface brightness are precisely measured.

M33 is one of the fewest objects for which it is possible to combine a high quality 
and high resolution RC from CO J=1-0  line that extends inward to 200~pc with a 
high quality 21-cm line RC which extends out to 13 disc scalelengths 
i.e. 19~kpc (Corbelli $\&$ Salucci 2000, Corbelli 2003). The galaxy 
has no prominent bar or bulge and the HI and CO velocity fields are very 
regular and cannot be explained by Newtonian dynamic without including  
a massive dark matter halo. 
M31 is a very large nearby galaxy whose extended and spatially resolved RC  
has been recently studied, in the Newtonian framework, to advocate a dark matter 
component contributing more than half the mass inside 30 kpc
(Widrow, Perrett $\&$ Suyu 2003, Geehan et al. 2006, Carignan et al. 2006). 

In this paper we use recent kinematical data of  these two galaxies to test 
MOND theory.  It is important to investigate the uncertainties left in the 
kinematics or mass distribution in order to claim complete failure or success 
of this theory. To this purpose we investigate M33 in Section 2,
and in Section 3 we check the compatibility of MOND with 
less probable but still possible deconvolution models for the M33 21-cm datacube.
In Section 4 we analyze MOND mass models for M31 using the most recent RC data.
Section 5 summarizes our results and addresses future observational strategies.

\section{The M33 rotation curve in the MOND framework}

Neutral gas in M33 extends well beyond the optical disc. The rotation curve 
and the average radial distribution of gas can be derived from 21-cm spectral 
line data using a set of tilted concentric rings that accounts for the possible 
presence of a warp. This is especially important when testing MOND because 
the kinematics and the distribution of matter in the outermost regions of 
galaxies are strongly coupled by the local non-Newtonian gravitational field.
    
The RC that we use is shown in Figure 1 with 2$\sigma$ errorbars in each bin 
derived from the dispersion of single spectra velocities after deconvolution.
Data have been deconvolved using the tilted ring model which best 
fits the Arecibo 21-cm data (hereafter deconvolution Model 1,
see bottom panels of Figure 1;  Corbelli $\&$ Schneider 1997,  
Corbelli $\&$ Salucci 2000). Uncertainties in this model will be investigated 
in the next Section.  The inner RC, over the interval $0.2 < R <3.4$~kpc, is  
from CO J=1-0 data at an angular resolution of 0.75~arcmin (Corbelli 2003). 
For $3.4\le R < 5.5$~kpc we use 21-cm interferometric observations of Newton 
(1980) at 1.5$\times$3 arcmin angular resolution. The outer RC is from Arecibo 
21-cm data at 4.5~arcmin angular resolution (Corbelli $\&$ Schneider 1997).
Corbelli $\&$ Salucci (2000) and Corbelli (2003) have used these data for modeling 
the M33 mass distribution when a dark matter halo is in place. Note however that
the distance assumed in Corbelli (2003)  is 760~kpc and not 840~kpc as erroneously
stated in that paper. Here we shall use a distance to M33 of 840~kpc (Freeman et 
al. 2001). As shown by Corbelli $\&$ Salucci (2000) the rotation curves of the 
receding and approaching sides of M33 look very similar for deconvolution Model 1. 
Their slopes are consistent (within 1$\sigma$ errors) with those relative to the 
rotation curve of the galaxy as a whole.

The visible mass components in M33 are: stars, distributed in a disc and in a 
spheroid, and gas, in molecular and in neutral atomic form. In the stellar disc we 
assume that the mass follows the light distribution in the K-band that is  well
fitted by an exponential law  with scalelength $R_d=5.8\pm 0.4$~arcmin 
($1.4\pm 0.1$~kpc) (Regan $\&$ Vogel 1994). There are no wiggles or deviations 
from a pure exponential disc between 3 and 18~arcmin. Beyond 18~arcmin the the 
near-infrared brightness drops below the sensitivity limit of the quoted 
observations. However preliminary work in the I-band by Ferguson et al. (2006) 
shows that the exponential surface brightness holds out to about 40~arcmin. 
In this Section we extrapolate such exponential stellar light 
distribution beyond 40~arcmin, to the outermost RC data point.
Alternative models are discussed in Section 3. For consistency with other works  
on MOND we assume an infinitely thin stellar disc. In the innermost region 
($R<3$~arcmin) there is a luminosity excess with respect to the exponential disc
extrapolation and a  small spheroid is likely to be in place
though its luminosity and mass are very uncertain (Regan $\&$ Vogel 1994; 
Corbelli 2003). We will take this stellar  component into account by  
parameterizing its contribution to the circular velocity as:

$$V_{sph} = G M_{sph}  R^{0.5}/(R+s)
\eqno(1) $$

\noindent
the total mass of the spheroid, $M_{sph}$, and its typical radius,
$s$, are free parameters. Observations suggest values of $s$ in the 0.1-1~kpc range
and spheroidal masses smaller than 10$\%$ of the stellar disc mass 
(see Corbelli 2003 and references therein). Being M33 a blue galaxy, also in
its central region, the only restriction we place on the spheroid here is 
$M_{sph}/L_{sph}<3$ where $L_{sph}=4\times 10^8$~$L_\odot$ is the largest blue 
spheroidal luminosity ever estimated. 
The atomic and molecular gas surface densities in M33 are shown by Corbelli (2003). 
They result from azimuthal averages of the data after deconvolution according to the
tilted ring model which best fits the 21-cm datacube.
The molecular mass of M33 is less than 10$\%$ the HI mass using the CO  
to H$_2$ conversion factor determined by Wilson (1995) in M33.  

The free parameters of the M33 mass model in the MOND framework  
are: $s$, $M_{sph}$, and $M_d/L_d$, (where $L_d=5.7\times10^9$~$L_\odot$ is the 
blue disc luminosity). We use the critical acceleration value $a_0$   
derived from the analysis of a sample of rotation curves  
$a_0 = 1.2\times 10^{-8}$~cm~s$^{-2}$ (Sanders $\&$ McGaugh 2002).
 
\begin{figure}
\includegraphics[width=84mm]{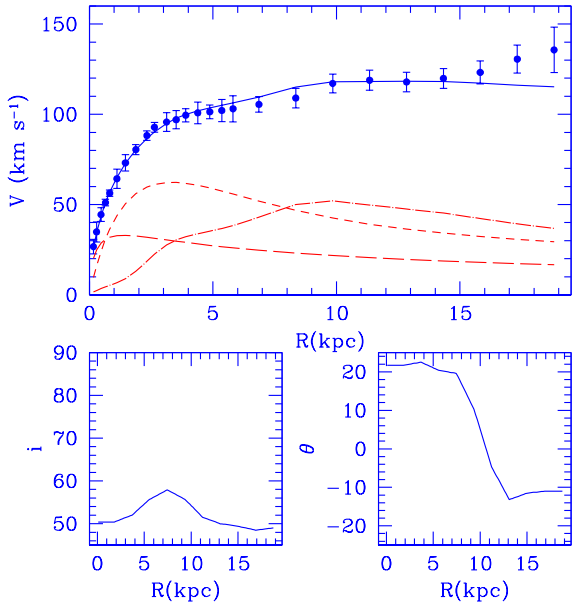}
\caption{Data for M33 rotation curve from deconvolution Model 1 and the MOND best
fit. The long and short dashed lines are the Newtonian RC of
the spheroid and stellar disk respectively. The dot-dashed line is the gas
Newtonian RC. The bottom panels
show the inclination $i$ and position angle $\theta$ of Model 1.}
\label{fig:Fig1}
\end{figure}

We use the reduced chi-square statistic, 
$\chi^2_\nu$, to judge the goodness of a model fit.
Figure 1 shows the MOND best-fit model curve: its $\chi^2_\nu = 2.0$  
is higher than the value $0.7$ obtained for dark halo models (Corbelli 2003).
The best fitting values of the free parameters are $M_d/L_d = 0.6 $, $s=1.4$~kpc,   
and $M_{sph}=1.2\times 10^9$~M$_\odot$ (the maximum allowed value).  
The fit is poor and only marginally acceptable, being 99.9$\%$ the probability
of finding $\chi^2_\nu \ge 2.2$ for random events. The rotational velocities 
predicted by MOND are higher than observed for $R\simeq 7-8$~kpc and lower 
than data for $R>15$~kpc. MOND has difficulties to reproduce the  M33 RC  because  
of the rising rotational velocities in the outer disc where the  
HI gas mass density strongly declines. Moreover, the resulting spheroidal mass, 
about 20$\%$ of the total stellar mass, is unrealistic for this
blue galaxy, it bounces to the extreme value allowed by our fit and its typical  
radius is larger than observational estimates. 
Variations of the stellar disc scalelength within the uncertainties quoted by
Regan $\&$ Vogel (1994) do not affect the above conclusions significantly. 

In modeling the gas distribution in the outer disc we have assumed that most 
of the gas is in neutral form and is traced by the 21-cm line. Absorption measurements 
of the 21-cm radiation from background sources behind the M33 disc have excluded 
the possibility that there are relevant HI masses in the outer disc of M33 which
are undetected in emission at 21-cm (Corbelli $\&$ Salpeter 1993a). There is 
however the possibility that extragalactic UV radiation ionizes 
the outer disc gas as the HI column density approaches the value of  
$10^{19}$~cm$^{-2}$ (Corbelli $\&$ Salpeter 1993b). This implies a sharp HI radial 
fall-off, as that observed along the M33 major axis by Corbelli, Schneider $\&$ Salpeter 
(1989). The gas extends beyond 18~kpc but it is undetected via 21-cm radiation because 
it is fully ionized. The sharpness of the HI edge is a function of the gravity perpendicular
to the disc and it is worth noticing that MOND models reproduce the observations as 
the Newtonian dynamic does in the presence of a dark halo.
The presence of an ionized component extending radially further out would rise 
the gas RC by 1-2 km $s^{-1}$ at the  outermost sampled radius. This is however
not sufficient to improve the significance of MOND fit to the RC.

\section{Can observations and MOND predictions in M33 reach a better agreement?}

In this Section we investigate whether MOND models can provide a better fit to
the M33 RC when uncertainties in the modeling presented in Section 2 are taken into 
account.

\noindent
{\it (a)Distance uncertainties}

For M33 we are using a distance of $D=840$~kpc  determined
from Hubble Space Telescope Cepheid data (Freeman et al. 2001). The 
uncertainties on this measured value are $\pm 40$~kpc. To be conservative 
we take into account other values obtained by different distance indicators. These fall 
in the range 790-940~kpc (McConnachie et al. 2004; McConnachie et al. 
2005 and reference therein; Bonanos et al. 2006; Sarajedini et al. 2006).
We model fit the M33 RC by assuming the distance as an additional free parameter.
The best fit is obtained for a distance to M33 of 860~kpc, but the
improvement in term of the $\chi^2_\nu$ value is negligible (1$\%$).

\noindent
{\it (b)Departures of the stellar mass distribution from a simple exponential law}

Preliminary work by Ferguson et al. (2006) shows that 
the surface brightness of the M33 disc has no departures from a pure exponential
law out to about 40~arcmin. Between 40 and 60~arcmin,
only RGB star counts are available and their distribution can be fitted by a
steeper exponential law than that describing the K and I-band inner surface 
brightness.
If the distribution of RGB stars reflects the stellar mass distribution in the 
outer disc we have to modify the stellar mass exponential scalelength used in the 
previous Section at large radii. We shall use a scalelength of 1.4~kpc 
(Regal $\&$ Vogel 1993) for $R\le 10$~kpc and of 0.9~kpc for 
$10<R<20$~kpc. For sake of completeness we will consider also models with a larger  
stellar scalelength, up to 2~kpc, for $R>10$~kpc.

In any case (for both smaller and larger scalelength at large radii) 
variations of the best fitting parameters and $\chi^2_\nu$ are
hardly noticeable. 
This is because the change in scalelength takes place in a region
where the stellar contribution to circular velocity is not prominent
and radially declining.

\noindent
{\it (c)Tilted ring models}
 
The modest inclination of M33 with respect to our line of sight 
implies that small variations of the inclination angle give 
non negligible variations of the deprojected rotational velocity.
In this Section we check if tilted ring models, with less statistical  
relevance but still within the 95$\%$ confidence level for fitting the
21-cm Arecibo datacube,  yield a rotation curve in closer agreement with 
MOND predictions.

The tilted ring model we use has 11 free rings. The fitting program 
interpolates between the 11 free rings to find the parameters at each radius 
(see Corbelli $\&$ Schneider 1997). 
Here we vary the parameters of the tilted ring model
as follows. We fix the ring parameters to the best fitting values of Model 1  
for $R<6$~kpc, and in the interval  $6\le R\le 15$~kpc we vary the inclinations 
of the 5 free rings by $\pm 5\deg$ respect to Model 1 values. 
Each of these models is compatible with the 21-cm datacube and yields 
a rotation curve over $0.2\le R\le 15$~kpc that we fit using MOND prescription.

The use of deconvolution models whose ring inclinations are  
systematically lower than those of Model 1 for $10\le R\le 15$~kpc, results in a 
RC which is flatter than that of Figure 1 and in better agreement with MOND model.
We then consider for these 5 free rings the deconvolution model which gives the 
lowest $\chi^2_\nu$ ,for MOND fit out to $R=15$~kpc, and we vary the position angle  
and inclination of the two outermost free rings  by $\pm 5,10,15$ deg.
We choose a combination of position angles and inclinations of the last two 
rings, compatible with 21-cm data, whose associated RC has the lowest $\chi^2_\nu$ 
for MOND model fit ($\chi^2_\nu=0.84$).
Figure 3 shows the whole M33 rotation curve derived according to the above
prescriptions and the relative deconvolution model (hereafter deconvolution Model 2). 
 
\begin{figure}
\includegraphics[width=84mm]{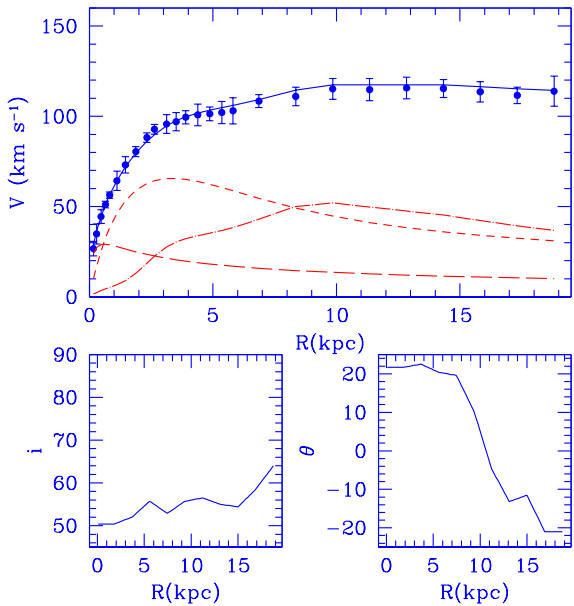}
\caption{Data for M33 rotation curve from deconvolution Model 2 and the MOND best
fit to it. The long and short dashed lines are the Newtonian RC of
the spheroid and stellar disk respectively. The dot-dashed line is the gas
Newtonian RC. The bottom panels show the inclination
$i$ and position angle $\theta$ relative to Model 2.}
\label{fig:Fig3}
\end{figure}

\begin{figure}
\includegraphics[width=84mm]{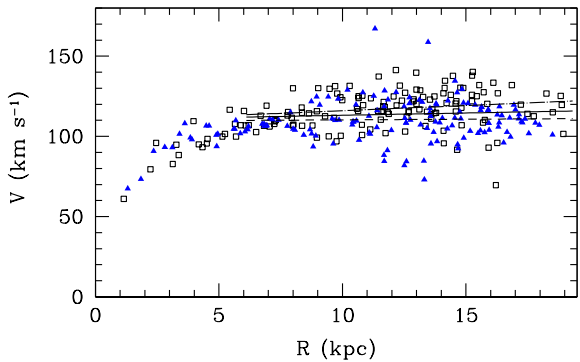}
\caption{Unbinned kinematical data from 21-cm line observed at Arecibo  using
 deconvolution Model 2. Filled triangles refer to the approaching side of the galaxy,
 and open squares to the receding side. Straight lines are the best linear fit for
 $R>6$~kpc to all data (continuous line), to the approaching side set (dashed line) 
 and to the receding side set (dashed-dotted line).}
\label{fig:Fig4}
\end{figure}

MOND fit  improves considerably when 
outermost orbits are more inclined and at more negative position angles
than in Model 1.
The velocity fields of the two separate halves of M33 look very similar 
also for Model 2 and are shown in Figure 3. The linear fit to the data 
for $R>6$~kpc have consistent slopes (within 1-$\sigma$ error). 
The parameters of the mass model
are $M_d/L_d = 0.7$, $s=0.6$~kpc  and $M_{sph}=4\times 10^8$~M$_\odot$.
The MOND best fit to Model 2 data is good and requires spheroidal parameters 
closer to values suggested by multiwavelengths observations of M33. 
Deconvolution Model 2, or similar ones, are therefore  acceptable
alternatives to Model 1 and give rotation curves compatible with MOND predictions
with high statistical significance, similar to dark matter halo models.
Only deeper 21-cm maps or H$\alpha$ emission searches around and beyond the
18~kpc radius can constrain the deconvolution model further and solve the
ambiguity left in M33 by the current data.

\section{The M31 rotation curve in the MOND framework}

Recent 21-cm observations of M31 along its major axis have shown that the 
rotation curve of this nearby galaxy stays flat out to 35~kpc 
(Carignan et al. 2006). Because of the
discrepancies between different determinations of the RC in the inner 10~kpc  
and the relative unimportance of this region for MOND tests,  fully into
the Newtonian regime, we  consider only the radial range and HI rotational velocities
given by Table 1 of Carignan et al. (2006).  For this radial range (10-35~kpc),
there are very few uncertainties on the galaxy inclination and position angle.
The warp of M31 is a minor one, due perhaps to the more massive and extended stellar
disc. Models of the warp at large radii imply a slightly increasing inclination 
as a function of radius (Newton $\&$ Emerson 1977; Briggs 1990). Due to the
nearly edge on position of the optical disc, the use of a constant inclination of
77$^o$ instead of a higher one at large $R$ overestimate the rotational
velocities by only 2-3$\%$. 21-cm maps suggest higher inclinations at large $R$ 
(Brinks $\&$ Burton 1984) but analysis of the extended stellar disc by Ibata et al. 
(2005) suggest instead a somewhat lower values of $i$.
The large errorbars in the rotation curve of this region take these uncertainties into 
account.

We consider a mass model with three 
components: a gaseous disc, a stellar disc and  a central spheroidal or
stellar bulge. This last component is very prominent and its
mass can be of the same order of the disc mass. Walterbos $\&$ Kennicutt (1987,1988) 
derive the bulge and disc optical luminosity profiles. There are very large 
uncertainties  on the bulge scalelength and on its mass because the disc-bulge 
decomposition is not very robust. The bulge scalelength range is $0.61\le r_b\le 1.8$~kpc 
(Geehan et al. 2006) and here we shall consider the two extremes of this range. For the 
disc scalelength we shall use that K-band photometric data which give a value of 4.5~kpc 
(Battaner et al. 1986). 
Similar values are found in the more extended R-band photometric maps (5.1~kpc out to 
40~kpc, Ibata et al. 2005).
The  bulge and disc blue luminosities are $9\times 10^9$~$L_\odot$
and $2\times 10^{10}$~$L_\odot$ respectively. The observed colors and 
population models imply stellar mass to light ratios between 2.8 and 6.5 in solar  
units (Bell $\&$ de Jong 2001) which we will use to limit our models.

We adopt the neutral gas surface density  measured by Sofue $\&$ Kato (1981).
The molecular gas mass is less than 10$\%$ of the gas mass (Nieten et al. 2006)
and its peak is located at about 11~kpc. Results are insensitive to the inclusion of 
this components as well as to the central supermassive black hole whose estimated mass 
is $\sim 0.5-1\times 10^8$~M$_\odot$ (Salow $\&$ Statler 2004; Bender et al. 2005).

\begin{figure}
\includegraphics[width=84mm]{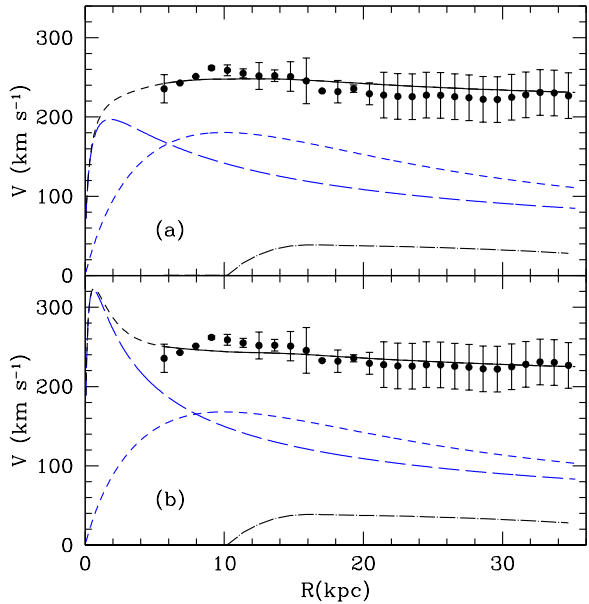}
\caption{Data for M31 rotation curve and the MOND best
fit to it. The long and short dashed lines are the Newtonian rotation curves of
the bulge and stellar disc respectively.The dot-dashed line is the gas 
Newtonian rotation curves. Panel $(a)$ shows the best MOND fit for
$r_b=0.61$~kpc, panel $(b)$ for $r_b=1.8$~kpc.}
\label{fig:Fig4}
\end{figure}

For $r_b=1.8$~kpc the minimum $\chi^2_\nu$ is 10.3 with $M_d/L_d=4.5$ and 
$M_b/L_b=6.5$.
For $r_b=0.61$~kpc the minimum $\chi^2_\nu$ is 8.1 with $M_d/L_d=3.9$ and 
$M_b/L_b=6.3$. MOND fits are shown in Figure 4.
The evident failure is due to the declining part of the rotation curve of M31 
between 10 and 20 kpc. While the classical Newtonian dynamic with a dark
matter halo is able to provide a good fit to the data (Widrow et al. 2003;
Carignan et al. 2006) by making the stellar disc responsible for the declining part
of the curve and the dark halo responsible for the outermost flattening,
MOND is  unable to reproduce this non monotonic behavior. It is in fact around
10~kpc that $g_n\sim a_0$ and non-Newtonian corrections start to be important and 
force a falling Newtonian RC into a flat one, inconsistent with the data.

Uncertainties on the M31 distance are much smaller than for M33.  
Distances resulting from different data sets and methods are always in
the range 770-795~kpc. Therefore possible variations with respect to the
assumed one (D=780~kpc) are less than 2$\%$ and give negligible effects on 
the MOND fit (Freeman et al. 2001; Joshi et al. 2003;
McConnachie et al. 2005 and reference therein; Ribas et al. 2005).
Concerning the uncertainties on the disc scalelength we have found no evidence 
for a scalelength smaller than indicated by K-band photometry. 
However from measurements of light profiles  in other bands, extending at
larger radii, we cannot exclude a flattening of the stellar light profile 
at large radii (Walterbos $\&$ Kennicutt 1988; Ibata et al. 2005; Irwin et 
al. 2005). MOND models which take into account a change in the disc scalelength 
at large radii do not give acceptable fits to M31 RC data.

\section{Conclusions} 

Local Group galaxies benefit from low observational uncertainties in distance
determination and in the distribution of baryonic matter. Moreover, they  provide 
RC data to an unbeatable spatial 
resolution. For this reason we have chosen the two nearest spiral galaxies, M31 
and M33, to test MOND. M33, a low luminosity late galaxy, has a very extended 
HI disc providing a rotation curve out to 13 disc scalelengths. Dark matter
dominates over the visible baryonic matter well inside its optical radius
according to Newtonian dynamic. However its inclination is not sufficiently high 
to limit the uncertainties on the rotational velocities of the outer disc. 
Here the presence of a warp leaves some degree of freedom to the tilted
rings deconvolution model. Even though MOND models are marginally compatible with
the rotation curve from the best fitting tilted ring model to the 21-cm datacube, 
data leave some space for a less rising curve, well fitted by MOND models. 
Since deconvolution models in better agreement with MOND theory predict a more 
inclined outermost disc,
future observations should  check whether this is effectively the case
by searching for faint emission above the main disc at a specific 
velocity.
Sensitive searches for extended emission should be carried out both at 21-cm 
and in the H$\alpha$ line (since there is the possibility that extragalactic 
UV radiation ionizes the outermost disc). 

In M31 the  disagreement between the RC and MOND predictions 
is statistically significant. MOND models are unable to reproduce the non 
monotonic behavior of the rotation curve in regions where
$g_n\sim a_0$ and non-Newtonian corrections predict a flat RC. However
the actual RC is made up of data with small uncertainties in some radial 
range while in others and beyond 20~kpc uncertainties 
are large. For this galaxy it would be nice to have more uniform  
uncertainties by producing a fully sensitive two dimensional map at 21-cm
which provides kinematical information on a more extended area and defines 
the orientation of the possible warp. 

We are grateful to the anonymous referee for her/his useful criticism to
the original manuscript which have improved the quality of the paper.

\label{lastpage}
\end{document}